\documentclass{article}
\usepackage{epsfig}

\date{\today}

\usepackage[intlimits,tbtags]{amsmath} 
\usepackage{amsfonts,amssymb}
\usepackage{amsthm}
\usepackage{amscd}

\begin{document}
\title{Holographic superfluids as duals of rotating black strings}
\author{{\large Yves Brihaye \footnote{email: yves.brihaye@umons.ac.be} }$^{\ddagger}$ and 
{\large Betti Hartmann \footnote{email: b.hartmann@jacobs-university.de}}$^{\dagger}$
\\ \\
$^{\ddagger}${\small Physique-Math\'ematique, Universite de
Mons-Hainaut, 7000 Mons, Belgium}\\ 
$^{\dagger}${\small School of Engineering and Science, Jacobs University Bremen, 28759 Bremen, Germany}  }

\date{}
\newcommand{\dd}{\mbox{d}}
\newcommand{\tr}{\mbox{tr}}
\newcommand{\la}{\lambda}
\newcommand{\ka}{\kappa}
\newcommand{\f}{\phi}
\newcommand{\vf}{\varphi}
\newcommand{\F}{\Phi}
\newcommand{\al}{\alpha}
\newcommand{\ga}{\gamma}
\newcommand{\de}{\delta}
\newcommand{\si}{\sigma}
\newcommand{\bomega}{\mbox{\boldmath $\omega$}}
\newcommand{\bsi}{\mbox{\boldmath $\sigma$}}
\newcommand{\bchi}{\mbox{\boldmath $\chi$}}
\newcommand{\bal}{\mbox{\boldmath $\alpha$}}
\newcommand{\bpsi}{\mbox{\boldmath $\psi$}}
\newcommand{\brho}{\mbox{\boldmath $\varrho$}}
\newcommand{\beps}{\mbox{\boldmath $\varepsilon$}}
\newcommand{\bxi}{\mbox{\boldmath $\xi$}}
\newcommand{\bbeta}{\mbox{\boldmath $\beta$}}
\newcommand{\ee}{\end{equation}}
\newcommand{\eea}{\end{eqnarray}}
\newcommand{\be}{\begin{equation}}
\newcommand{\bea}{\begin{eqnarray}}

\newcommand{\ii}{\mbox{i}}
\newcommand{\e}{\mbox{e}}
\newcommand{\pa}{\partial}
\newcommand{\Om}{\Omega}
\newcommand{\vep}{\varepsilon}
\newcommand{\bfph}{{\bf \phi}}
\newcommand{\lm}{\lambda}
\def\theequation{\arabic{equation}}
\renewcommand{\thefootnote}{\fnsymbol{footnote}}
\newcommand{\re}[1]{(\ref{#1})}
\newcommand{\R}{{\rm I \hspace{-0.52ex} R}}
\newcommand{\N}{{\sf N\hspace*{-1.0ex}\rule{0.15ex}%
{1.3ex}\hspace*{1.0ex}}}
\newcommand{\Q}{{\sf Q\hspace*{-1.1ex}\rule{0.15ex}%
{1.5ex}\hspace*{1.1ex}}}
\newcommand{\C}{{\sf C\hspace*{-0.9ex}\rule{0.15ex}%
{1.3ex}\hspace*{0.9ex}}}
\newcommand{\eins}{1\hspace{-0.56ex}{\rm I}}
\renewcommand{\thefootnote}{\arabic{footnote}}

\maketitle

\bigskip

\begin{abstract}
We study the breaking of an Abelian symmetry close to the horizon of
an uncharged rotating Anti--de Sitter black string in 3+1 dimensions.  
The boundary theory living on $\mathbb{R}^2\times S^1$ has no rotation, but a magnetic field
that is aligned with the axis of the black string. This boundary theory describes non--rotating
(2+1)--dimensional holographic superfluids with non--vanishing superfluid velocity.
We study these superfluids in the grand canonical ensemble and show that for sufficiently
small angular momentum of the dual black string and sufficiently small superfluid
velocity the phase transition is 2nd order, while it becomes 1st order
for larger superfluid velocity. Moreover, we observe that 
the phase transition is {\it always} 1st order above a critical value of the angular momentum
independent of the choice of the superfluid velocity.
\end{abstract}
\medskip
\medskip

\section{Introduction}
The gravity/gauge theory duality \cite{ggdual} has attracted a lot of attention in the past years. The most famous
example is the AdS/CFT correspondence \cite{adscft} which states that a gravity theory in a $d$-dimensional
Anti--de Sitter (AdS) space--time is equivalent to a Conformal Field Theory (CFT) on the $(d-1)$-dimensional boundary of AdS.

Recently, this theory has been used to describe strongly interacting systems in terms of gravity duals.
In particular, so-called holographic superconductors and holographic superfluids can be described with the help of black holes in
higher dimensional space--time \cite{gubser,hhh,reviews}.  The main idea is that close to the horizon of a black hole in AdS space--time the Abelian symmetry can be spontaneously broken since the charged scalar field acquires a mass.
The reason for this is that
close to the horizon of the black hole the effective mass of the scalar field can become
negative with masses below the Breitenlohner--Freedman bound \cite{bf} such that the scalar
field becomes unstable and possesses a non--vanishing value on and close to the horizon
of the black hole. The local bulk U(1) symmetry  is associated to a global U(1) symmetry on the AdS boundary and
the value of the scalar field on the AdS boundary  with the corresponding condensate in the dual theory.
The Hawking temperature of the black hole is identified with the temperature of the dual theory. Below a certain
critical temperature the black hole becomes unstable to form scalar hair which in the dual theory
is interpreted as the onset of superconductivity. One of the most striking properties of
superconductors is the Meissner--effect, i.e. the expulsion of a magnetic field from
a superconductor and the transition from type--I to type--II superconductors. Consequently magnetic fields
have been considered in the study of holographic superconductors \cite{hhh,nakano,johnson,aekk,mps,gwwy,hks,mno}.
In \cite{hks} the non---vanishing spatial components of the U(1) gauge field on the AdS boundary have been
interpreted as non--vanishing superfluid currents in the description of holographic superfluidity. It was
shown that for sufficiently large superfluid velocity the phase transition changes from being 2nd to 1st order. Interestingly, this changes when taking strong back--reaction of the space--time into account. In \cite{sonnerwithers} it was shown that in this case the phase transition remains
2nd order for all values of the superfluid velocity.

In most studies the AdS black holes are static and the dual theory hence describes non--rotating
superconductors. In \cite{sonner} a Kerr--Newman--AdS black hole has been considered in order to describe (2+1)--dimensional 
rotating superconductors living on the surface of a sphere and it was shown that a critical value of
the rotation parameter exists such that superconductivity gets destroyed. This was interpreted as being analogous
to a critical magnetic field.

In this paper we are studying symmetry breaking of an Abelian symmetry close to the
horizon of an uncharged rotating black string in (3+1)--dimensional AdS space--time.
Uncharged  black string solutions have first been discussed in \cite{lemos1,lemos2}, while the charged generalizations have been given in \cite{lemos3,cai}. The thermodynamics of the latter
have also been considered \cite{dehghani3}, while a generalization including a dilaton field has been constructed in \cite{dehghani2}. In
\cite{dehghani} the Abelian--Higgs model with a Mexican-hat potential has been studied in the background
of charged rotating black strings and it was shown that the black strings can be dressed
by Abelian--Higgs hair providing a further counterexample to the No--hair conjecture. This conjecture states that a stationary black hole in 4--dimensional space--time is uniquely characterized by a 
finite number of quantities that are subject to a Gauss law \cite{nohair}.
These are the black hole's mass and angular momentum if the black hole is uncharged and the mass, charge and angular momentum if it is charged.

Though the boundary theory that lives on $\mathbb{R}^2\times S^1$ (i.e. on the surface
of a cylinder) has no rotation there is a magnetic field on the boundary aligned with the axis
of the black string. In the gauge/gravity duality we interpret the boundary theory
to describe a non--rotating superfluid with non--vanishing superfluid currents. 
Our work can be seen from two different viewpoints  which we will both take in this paper.
Looking at the gravity side, our model is a modification of the model studied in \cite{dehghani} in the sense that we replace the Mexican-hat potential by a potential of the form $m^2\vert\Psi\vert^2$
and that we choose as background metric that of an uncharged black string.
We will then study the formation of scalar hair on the uncharged black string for this potential.
Looking at the boundary theory, we are discussing an extension of the work done in \cite{hks}. For
vanishing angular momentum, our model will be equivalent to that used in \cite{hks} and
various other studies on holographic superconductors and superfluids in 2+1 dimensions with the
difference that in this paper the spatial boundary is $\mathbb{R}\times S^1$, while
in \cite{hks} it is $\mathbb{R}^2$. In particular, we will see that our results
for vanishing angular momentum agree with those in \cite{hks}.

Our paper is organised as follows: in Section 2, we give the model including the equations
and boundary conditions. We also discuss the free energy in this section.
 In Section 3, we discuss our numerical results. This section
is subdivided into two parts: in the first part, we present our results from the gravity
perspective, in particular we will fix the value of the horizon to a constant value.
In the second part of section 3, we give our results for the boundary theory describing
holographic superfluids. In this latter case we allow the horizon radius and hence the temperature of
the black string to vary. We give our Conclusions in Section 4.

\section{The Model}
In this paper, we are studying the formation of scalar hair on a rotating black string in $(3+1)$--dimensional Anti--de Sitter space--time. 
The matter action of a complex scalar field minimally coupled to a U(1) gauge field reads~:
\begin{equation}
\label{action}
S= \int d^4 x \sqrt{-g} \left(-\frac{1}{4} F_{\mu\nu} F^{\mu\nu} + \left(D_{\mu}\Psi\right)^* D^{\mu}\Psi - m^2 \Psi^*\Psi \right) \ \ \ \ , \ \  \mu,\nu=0,1,2,3  \ ,
\end{equation}
where $F_{\mu\nu} =\partial_{\mu} A_{\nu} - \partial_{\nu} A_{\mu}$ is the field strength tensor and
$D_{\mu}\Psi=\partial_{\mu} \Psi - ie A_{\mu} \Psi$ is the covariant derivative.
$e$ and $m^2$ denote the gauge coupling and the mass of the scalar field $\Psi$, respectively.
The equations of motion resulting for the variation of this action read
\begin{eqnarray}
 \frac{1}{\sqrt{g}} D_{\mu}\left(\sqrt{-g}g^{\mu\nu} D_{\nu}\Psi\right) &=& m^2 \Psi \ \ , \\
 \frac{1}{\sqrt{g}} \partial_{\mu}\left(\sqrt{-g}g^{\mu\nu} g^{\rho\sigma} F_{\nu\sigma}\right)&=&e g^{\rho\lambda} J_{\lambda} \ ,
\end{eqnarray}
where $J_{\lambda}=i\left[\Psi^* \left(D_{\lambda}\Psi\right) - \Psi \left(D_{\lambda} \Psi\right)^* \right]$ is the
4--current. 

From the viewpoint of the dual theory we are working in the so--called probe limit in the following. On the gravity side this means that we are studying (\ref{action}) in the background of a fixed space--time. Here, this space--time is that of a rotating
black string with metric \cite{lemos1,lemos2}
\be
ds^2 = N(r) \left(dt+ L(r) d\varphi\right)^2 - K(r) d\varphi^2 - \frac{1}{f(r)} dr^2 - \frac{r^2}{\ell^2} dz^2  \ , 
\ee 
where 
\be
     N(r)= f(r) \xi^2 - \frac{r^2}{\ell^4}a^2 \ \ , \ \ f(r)  = - \frac{\ell M}{r} + \frac{r^2}{\ell^2} \ \ ,  \ \ L(r) = a \xi \frac{M \ell}{rN(r)}  \ \ , \ \ K(r) = N(r) (L(r))^2 - f(r) a^2 + r^2 \xi^2
\ee
with $\xi = \sqrt{1+a^2/\ell^2}$. $a$ is the rotation parameter and $\Lambda = -3/\ell^2$ is the cosmological constant. The mass ${\cal M}$ and the angular momentum ${\cal J}$ of our solution is given by \cite{dehghani3}~:
\begin{equation}
 {\cal M}=\frac{1}{8}(3\xi^2 -1)M \ \ , \ \ {\cal J}=\frac{3}{8}\xi M a \ ,
\end{equation}
such that the angular momentum ${\cal J}$ is a monotonically increasing function of the parameter $a$.
 
For $a=0$ we have non--rotating black strings with $N=f$, $L=0$, $K = r^2$. These solutions
are equivalent to the planar black hole solutions frequently used in the study
of holographic superconductors and superfluids in 2+1 dimensions. The only difference
is that for the planar black holes the boundary theory lives on $\mathbb{R}^3$, while
here it lives on $\mathbb{R}^2\times S^1$. Hence one dimension is compactified.

The horizon of the black string is at $r_h=\ell \sqrt[3]{M}$, while the surface
with $r_c=r_h\sqrt[3]{1+a^2/\ell^2}>r_h$ is the static limit. Since the black string is rotating
around the $z$-axis, we will have frame dragging effects. The angular velocity (of the horizon)
is given by
\begin{equation}
 \Omega_{r_h}=\frac{a}{\xi \ell^2}  \ .
\end{equation}
Note that for $r\rightarrow \infty$, the space--time reduces to ordinary AdS such that
the boundary theory has no rotation.
In the gravity/gauge duality the temperature of the boundary theory is given by the 
Hawking temperature $T$ of the black hole which reads \cite{dehghani3} 
\be
  T = \frac{f'(r_h)}{4 \pi \xi} = \frac{3 r_h}{4 \pi \xi \ell^2}  \ .
\ee
\subsection{Ansatz and equations of motion}
We use the following Ansatz for the gauge field $A_{\mu}$ and complex scalar field $\Psi$
\be
    A_{\mu}dx^{\mu} = \phi(r) dt + A(r) d\varphi \ \ \ , \ \ \Psi= e^{-i\omega t + i n \varphi} \psi(r)  \ \ , 
\ee
where $n\in {\mathbb{Z}}$ and $\omega$ is a constant.
Using the gauge transformation $\Psi\rightarrow \Psi e^{i\chi}$, $A_{\mu}\rightarrow A_{\mu} + \partial_{\mu}\chi$
with $\chi=i\omega t-in\varphi$ the scalar field becomes real. 
The equations of motion then read  
\be
\label{phi_eq}
   \phi'' - \frac{1}{2}\left(\frac{N'}{N} - \frac{K'}{K} - \frac{f'}{f} - \frac{2}{r} + \frac{L'L N}{K}\right) \phi' 
   - e^2 \phi \frac{1}{f}  \psi^2 + A' \frac{L' N}{K}  = 0  \ ,
\ee
\be
\label{a_eq}
    A'' + \frac{1}{2}\left( \frac{N'}{N} - \frac{K'}{K} + \frac{f'}{f} + \frac{2}{r} + \frac{2L'L N}{K}\right)A'
    - e^2\frac{1}{f}A \psi^2 - \phi'\left(\frac{LN'}{N} + L' - \frac{LK'}{K} + \frac{L'L^2 N}{K} \right)=0
\ee
for the gauge field functions and 
\be
\label{psi_eq}
   \psi'' + \frac{1}{2}\left(\frac{N'}{N} + \frac{K'}{K} + \frac{f'}{f} + \frac{2}{r}\right) \psi'
    - m^2 \frac{\psi}{f} 
    +e^2\left(\frac{\phi^2}{fN} - \frac{\left(L\phi  - A\right)^2}{fK}\right)  \psi=0
\ee
for the scalar field function. Here and in the following the prime denotes the derivative with respect to $r$. The non--vanishing components of the field strength tensor are $F_{rt}=\phi'$ and 
$F_{r\varphi}=A'$, which corresponds to a electric field in radial direction and a magnetic field
in $z$-direction, i.e. a magnetic field aligned with the axis of the black string.

Note that the equations are invariant under the rescalings
\be
      r \to \lambda r \ \ , \ \ \phi \to \lambda \phi \ \ , \ \ A \to \lambda A \ \ , 
      \ \ \psi \to \psi \ \ , \ \ M \to \lambda^3 M  \ 
\ee
with $\lambda$ an arbitrary constant. We will later use this rescaling to fix some of the parameters
of the model to particular values without loosing generality.

In general the set of coupled ordinary differential equations (\ref{phi_eq})-(\ref{psi_eq}) cannot
be solved analytically, however for $\psi\equiv 0$  there is a family of explicit  solutions~:
\be
\label{maxwell}
\phi(r)= c_1 - \frac{q}{r} \ \ , \ \  A(r) = c_2 + \frac{aq}{\xi r}
\ee
where $c_1$, $c_2$, $q$ are arbitrary constants.

Now and in the following, we will choose $m^2=-2/\ell^2$ which is larger than the Breitenlohner--Freedman bound $m_{\rm BF}^2=-9/(4\ell^2)$ \cite{bf}. 
The effective mass of the scalar field can then be read off from (\ref{psi_eq}) and is given by~:
\begin{equation}
 m_{\rm eff}^2 = m^2 +  e^2 \left(-\frac{\phi^2}{N}+ \frac{\left(L\phi - A\right)^2}{K}\right) = -\frac{2}{\ell^2} +  e^2 \left(-\frac{\phi^2}{N}+ \frac{\left(L\phi - A\right)^2}{K}\right)
\end{equation}
If the term in the brackets becomes sufficiently negative then we find that $m_{\rm eff}^2 < m_{\rm BF}^2$ and the scalar field
will become unstable. This will lead to a symmetry breaking and the formation of scalar hair on the black string.
For $r_h < r < r_c$, the metric function $N(r)$ is negative, hence it is not the first
term in the brackets that is responsible for the symmetry breaking, but rather the
second term. Note that $K(r_h)=0$ and $K(r_h) < 0$ for $r\in ]r_h:r_c[$ such that
$K(r\rightarrow r_c)\rightarrow -\infty$ if $r < r_c$ and 
$K(r\rightarrow r_c)\rightarrow +\infty$ if $r > r_c$. Hence in the ergoregion
$ r_h < r < r_c$ it is the second term that leads to $m_{\rm eff}^2$ being less than $-2/\ell^2$, while for $r > r_c$ it is the first term.

\subsection{Boundary conditions}
Though $N(r_c)=0$ all equations are regular at $r=r_c$ and can therefore
 be integrated numerically for $r\in [r_h, \infty]$. 
Requiring that the matter fields are regular at the horizon $r_h$ we need to impose the following
 three independent conditions~:
\be
\label{regular}
    \phi(r_h) = - \left.\frac{a A}{\l^2 \xi}\right\vert_{r=r_h} \ \ , \ \ 
    \phi'(r_h) =  \left.\frac{A \ell^4 \psi^2 - 3 r_h A' (\ell^2 + a^2)}{3 a \xi r_h \ell^2}\right\vert_{r=r_h} \ \ , \ \ 
    \psi'(r_h) = \left.\psi \ell^2 \frac{m^2 r_h^2(\ell^2 + a^2) + A^2 \ell^2}{3 (\ell^2 + a^2) r_h^3}\right\vert_{r=r_h} \ .
\ee
On the AdS boundary the fields have the following behaviour
\be
\label{infty}
      \psi(r \gg 1) = \frac{\psi_-}{r} + \frac{\psi_+}{r^2} \ \ , \ \ 
      \phi(r \gg 1) = \mu - \frac{Q_e}{r} \ \ , \ \  
      A(r \gg 1) = \sigma - \frac{Q_m}{r} .
\ee
In the following we will choose $\psi_-\equiv 0$, i.e. using the gravity/gauge duality language we will study a scalar operator ${\cal O}$ of conformal
dimension two. The expectation value $<{\cal O}>\equiv \psi_+$ will then be interpreted as the
value of the condensate in the dual theory. Moreover, in the gravity/gauge duality
$\mu$ and $\sigma$ can be interpreted as chemical potential and superfluid velocity, respectively \cite{hks}, while $Q_e$ and $Q_m$ are the electric and magnetic charge, respectively.

Note that the gauge field functions $\phi(r)$ and $A(r)$ asymptotically behave like the exact solution (\ref{maxwell}).

\begin{figure}[ht]
\hbox to\linewidth{\hss%
	\resizebox{9cm}{7cm}{\includegraphics{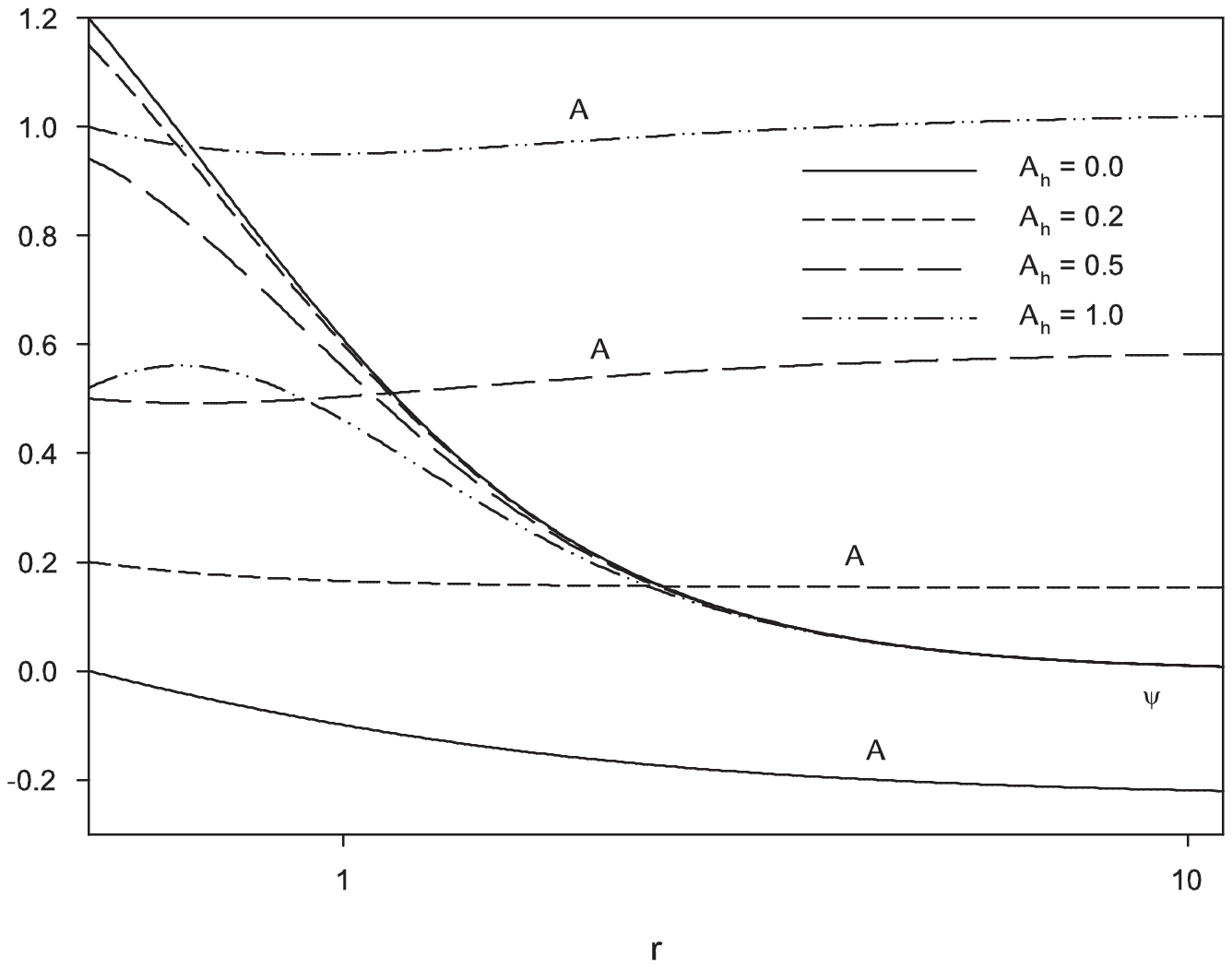}}
\hspace{5mm}%
        \resizebox{9cm}{7cm}{\includegraphics{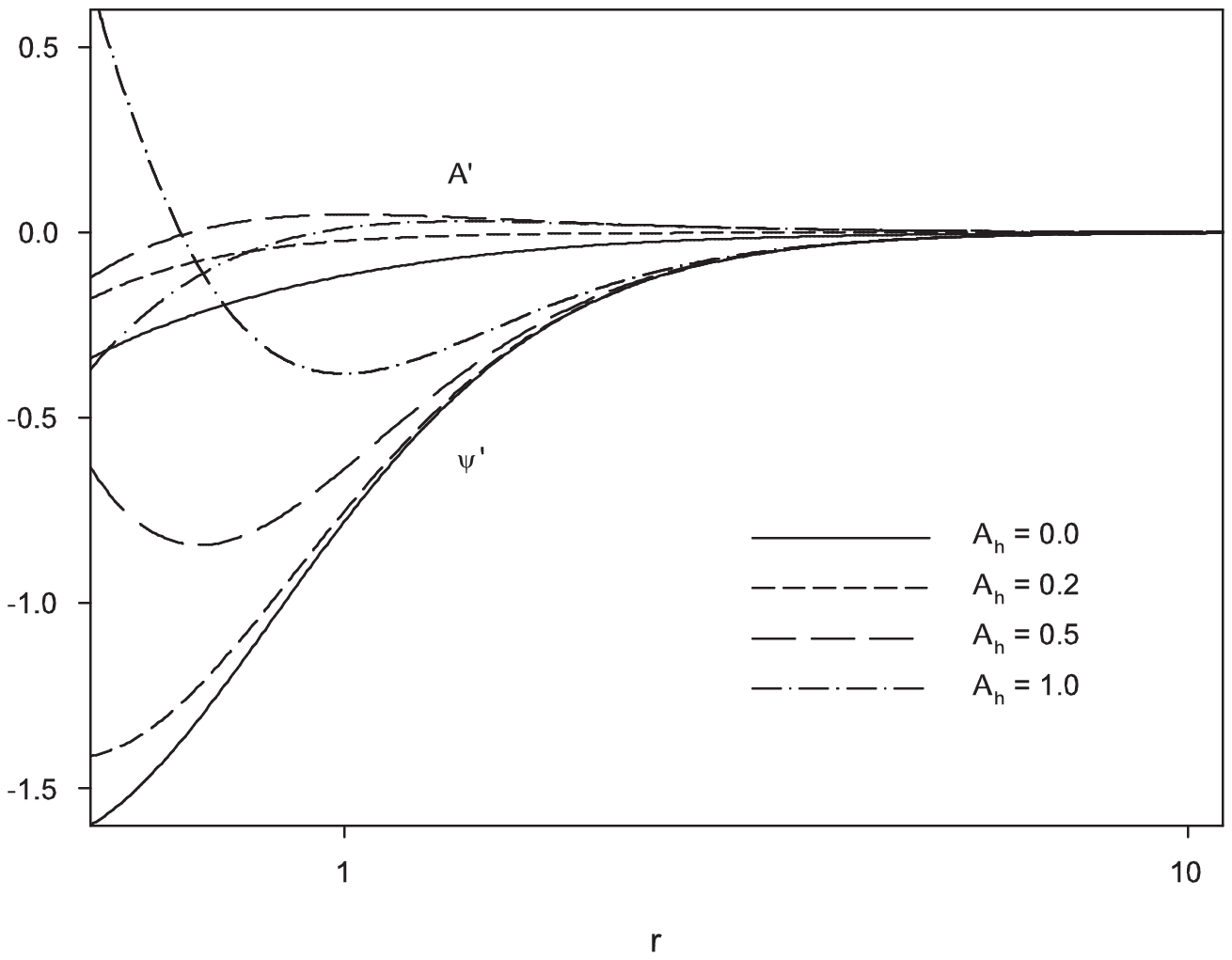}}	
\hss}
 
\caption{
{\small
The gauge field function $A(r)$ and the scalar field function  $\psi(r)$  are shown
for different values of $A(r_h)\equiv A_h$. Here $a=0.1$  and $\psi_+= 1.0$ (left). 
The derivative $A'(r)$ of the gauge field function $A(r)$ and the derivative $\psi'(r)$
of the scalar field function $\psi(r)$ for different values of $A(r_h)\equiv A_h$.
Here $a=0.1$  and $\psi_+= 1.0$ (right).
}
 }
\label{fig1}
\end{figure}
 \begin{figure}[ht]
\hbox to\linewidth{\hss%
	\resizebox{9cm}{7cm}{\includegraphics{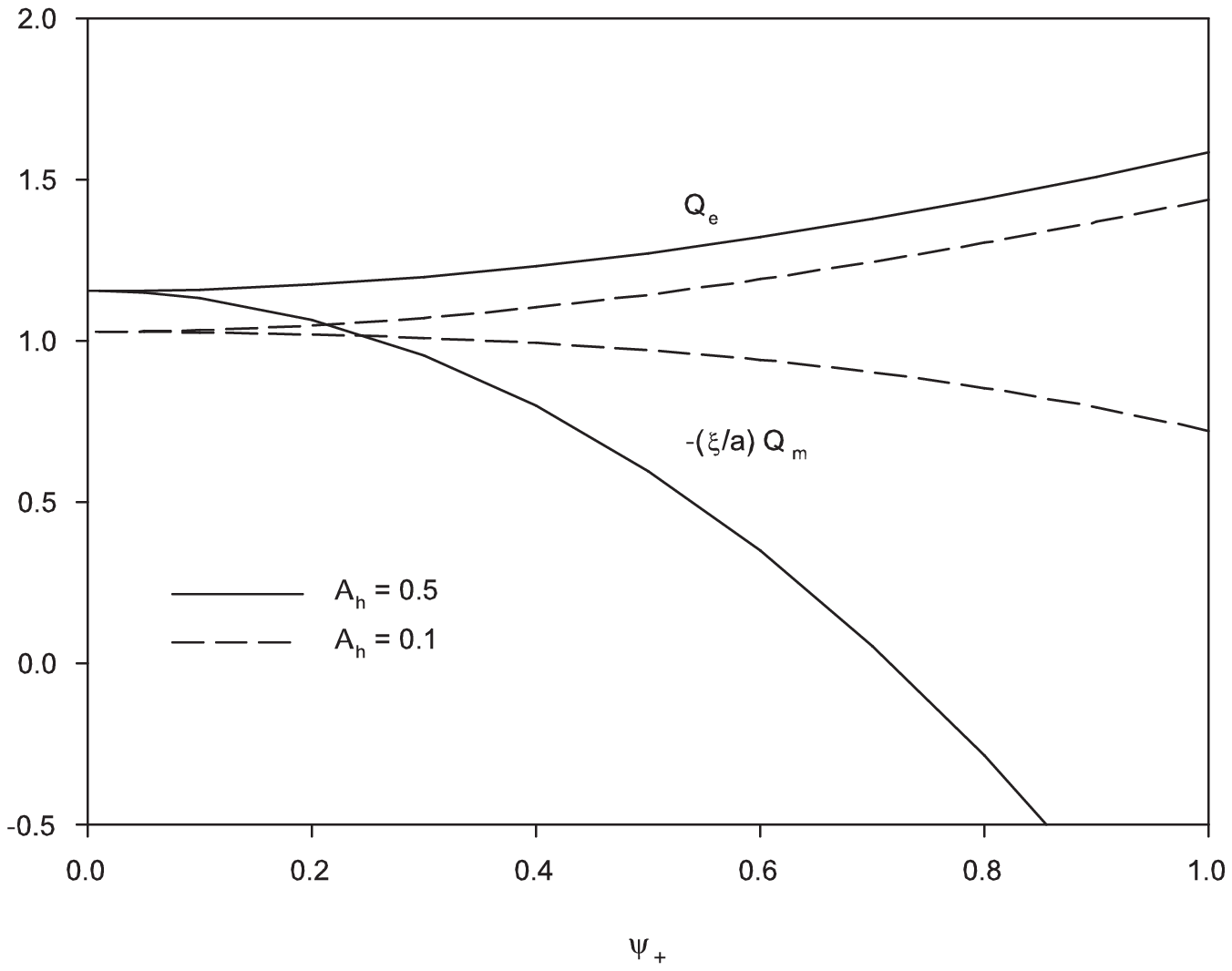}}
\hspace{5mm}%
        \resizebox{9cm}{7cm}{\includegraphics{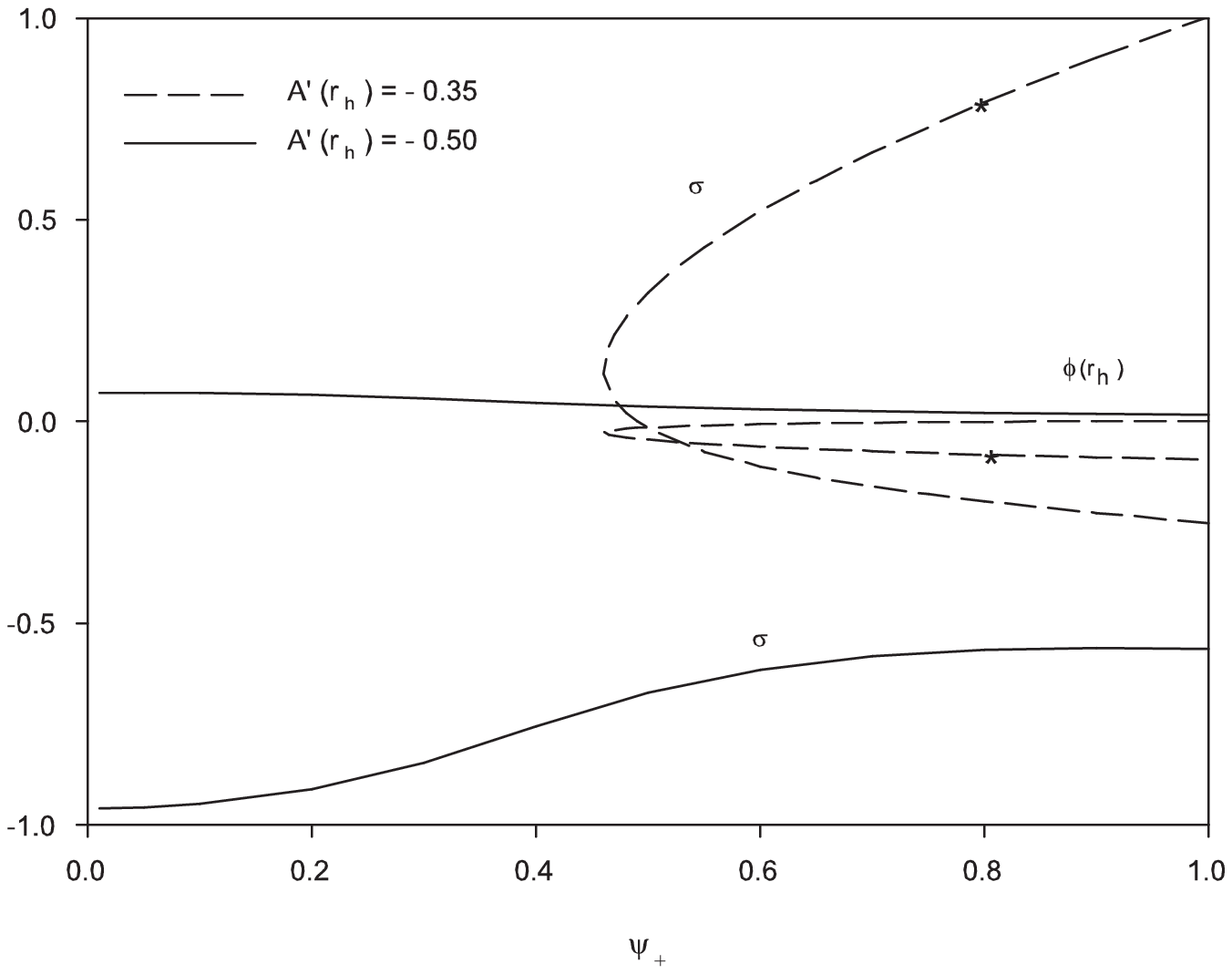}}	
\hss}
 
\caption{
{\small
The values of the electric charge $Q_e$ and of the magnetic charge $Q_m$ as function of
$\psi_+$  for $a=0.1$ and two different values of $A(r_h)\equiv A_h$ (left).
The values of $\sigma$ and $\phi(r_h)$ as function of $\psi_+$ for $a=0.1$ and
two different values of $A'(r_h)$. The star indicates the first branch of solutions (right).
}
 }
\label{fig2}
\end{figure}
\subsection{Free energy}

The free energy in the Grand Canonical Ensemble is given by $\Omega = - T S_{\rm os}$ \cite{hks}, where
$S_{\rm os}$ is the action (\ref{action}) evaluated on-shell. Using (\ref{action}) and the
equations of motion (\ref{phi_eq})-(\ref{psi_eq}) we find the following expression (with $\ell=1$ here and in the following)~:
\begin{eqnarray}
\label{a_onshell}
S_{\rm os}&=&\left.\int d^3 x\left[r^2\left(\frac{f(K-L^2N)}{2KN} \phi'\phi - \frac{f}{2K}A'A
+ \frac{fL}{4K} (A\phi' + A'\phi) - f \psi' \psi\right)\right]\right\vert_{\infty} \nonumber \\
&-& \int d^4 x \ r^2 \left(\frac{K-L^2N}{KN} \phi^2\psi^2 + 2 \frac{L}{K}A\phi\psi^2
- \frac{1}{K}A^2 \psi^2\right)   \ .                          
\end{eqnarray}
Note that in the limit of vanishing angular momentum $a=0$ expression (\ref{a_onshell}) reduces exactly
to that given in \cite{arean}. The boundary term in (\ref{a_onshell}) can be evaluated using
the behaviour of the matter functions at infinity (see (\ref{infty})). It is convenient 
to work with dimensionless variables and we thus study the dimensionless ratio $\Omega/(T^3 V)$, where
$V$ is the volume of the boundary. 
We then obtain
\begin{equation}
\label{onshell2}
\frac{\Omega}{T^3 V}=-\frac{1}{2}\left(Q_e \mu - Q_m \sigma\right)
+ \int\limits_{r_h}^{\infty} dr \ r^2 \left(\frac{K-L^2N}{KN} \phi^2\psi^2 + 2 \frac{L}{K}A\phi\psi^2
- \frac{1}{K}A^2 \psi^2\right)      \ .                       
\end{equation}
For the exact solution  (\ref{maxwell}) the integral term vanishes and the
free energy is simply given by the boundary term. 
For $\psi\neq 0$ the integral has to be computed numerically. We present our results
in the Numerical results section below. However, we will follow the analytic treatment done in \cite{abk} to get an
idea on how the free energy behaves. For that we first rewrite the free energy with the help
of the equations of motion as follows
\begin{equation}
 \label{free_energy1}
\frac{\Omega}{T^3 V}=\int\limits_{r_h}^{\infty} dr \ r^2 \left(\frac{f(L^2 N-K)}{2KN} \phi'^2 + \frac{f}{2K} A'^2 - \frac{fL}{K} \phi' A'\right)  \ .
\end{equation}
Now, we consider a perturbation around the exact solution (\ref{maxwell}) that is of the 
form \cite{abk}~:
\begin{equation}
 \psi=\varepsilon \psi_0 + O(\varepsilon^2) \ \ \ , \ \ \ 
A=A_0 + \varepsilon^2 \delta A + O(\varepsilon^4) \ \ \ , \ \ \
\phi=\phi_0 + \varepsilon^2 \delta \phi + O(\varepsilon^4) \ \ \ ,
\end{equation}
where $\phi_0$ and $A_0$ denote the exact solution (\ref{maxwell}). Inserting this into (\ref{free_energy1}) and taking into account that the boundary values of $A$ and $\phi$ also
change with $O(\varepsilon^2)$ \cite{abk} we find
\begin{equation}
 \frac{\delta \Omega}{T^3 V} = \varepsilon^4\left[Q_m\delta\sigma -Q_e \delta\mu\right] -\varepsilon^4 \int\limits_{r_h}^{\infty} dr \ r^2 \left( \frac{f(K-L^2 N)}{2KN} (\delta \phi)'^2 - \frac{f}{2K} (\delta A)'^2 + \frac{fL}{K} (\delta\phi)' (\delta A)'\right) + O(\varepsilon^6) \ .
\end{equation}
Let us first summarize the results for $a=0$ such that $L=0$, $N=f$ and $K=r^2$. This case has been discussed in \cite{abk}.
For $A=0$ it is obvious that $\delta \Omega/(T^3 V)$ is negative signaling a 2nd
order phase transition. For $A\neq 0$, however, the free energy can become positive, which
would indicate a 1st order phase transition. 

Now let us discuss what happens for $a\neq 0$. In this case, we have $L\neq 0$ and necessarily $A\neq 0$. To understand the dependence on $a$ let us write the prefactors of the three terms 
under the integral explicitely. These are
\begin{equation}
 \frac{f(K-L^2N)}{KN} = \frac{a^2 M}{r^3} + 1 \ \ , \ \ 
\frac{f}{K}=-\frac{ M}{r^3} - \frac{a^2 M}{ r^3} + 1 \ \ , \ \ 
\frac{fL}{K} = \frac{a M}{r^3} \xi \ .
\end{equation}
The first and third expression are obviously positive, while the second expression is
negative at and close to the horizon $r_h$ with $(f/K)\vert_{r_h}=-a^2/\ell^4$, while it is positive
for large $r$. It turns out that both 
$(\delta\phi)'$ and $(\delta A)'$ increase in absolute value for increasing $a$. Hence,
$\frac{\delta \Omega}{T^3 V}$ can become positive when increasing $a$ for relatively
small value of $\sigma$ such that
the phase transition changes from being 2nd to 1st order. We will see that this is confirmed
by our numerical results.

\section{Numerical results}
This section is subdivided into two parts. The first part contains our results
on the formation of scalar hair on rotating black strings. Hence, in the first section
we are taking the gravity perspective and are following the studies in \cite{dehghani2,dehghani}.
The formation of scalar hair on black holes is by itself interesting since
it provides a further counterexample to the No--hair conjecture \cite{nohair}. While a
Mexican-hat type potential was used in \cite{dehghani2,dehghani} we use
the above mentioned $m^2\psi^2$--potential and in addition choose the black string to be uncharged.
The second part of this section discusses the boundary theory of our model which
describes (2+1)--dimensional holographic superfluids. 

\subsection{Formation of scalar hair on rotating black strings}

The parameters $r_h$, $\ell$ and $a$ determine the space--time geometry. We will fix these and then
construct branches of black string solutions labelled e.g. by the values $\psi_+$ and $A(r_h)$. 
For our numerical calculations we have chosen $\ell=1$ and $r_h=0.5$.

Note that for $A(r_h)=0$ which implies $\phi(r_h)=0$ by the regularity condition (see (\ref{regular}))
we find that the gauge field functions $A(r)$ and $\phi(r)$ are proportional to each other~:
\be
\label{ah_zero}
          A(r) = - \frac{a}{\xi} \phi(r) \ .
\ee
The change of the gauge field function $A(r)$ and of the scalar field function $\psi(r)$ with changing $A(r_h)\equiv A_h$ is shown in Fig.\ref{fig1} (left) for
$a=0.1$ and $\psi_+=1.0$. The figure demonstrates in particular that
the scalar field function $\psi(r)$ develops a local maximum for sufficiently large values of the parameter $A_h$. We also give the $r$-derivatives of the gauge field function $A(r)$ and of the scalar field function $\psi(r)$ in Fig.\ref{fig1} (right).

To understand the pattern of solutions better we first fixed $A(r_h)\equiv A_h$ and varied
$\psi_+$. Our results are shown in Fig.\ref{fig2} (left) where we give the
electric charge $Q_e$ and the magnetic charge $Q_m$ in dependence on $\psi_+$ for
$a=0.1$ and two different values of $A_h$. We find that for both choices of $A_h$, the electric
charge $Q_e$ increases with $\psi_+$, while $Q_m$ decreases. At $\psi_+=0$ we find that $Q_e/Q_m=-\xi/a$.
In this limit the gauge field functions tend to the analytical solution (\ref{maxwell}). Note that
the value of $Q_e$ (and hence $Q_m$) depends on the choice of $A_h$. To state it differently: in the 
limit $\psi_+\rightarrow 0$ a continuum of solutions given by (\ref{maxwell}) exists. For fixed $A_h$ the solution
tends to a very specific value of the constants $c_1$, $c_2$ and $q$ in this limit.

We also constructed
families of black string solutions with fixed $A'(r_h)$. This seem somehow natural since the magnetic field
on the horizon is directly related to $A'(r_h)$. Our results are shown
in Fig.\ref{fig2} (right) where we show the value of $\sigma$ and of the gauge field
$\phi(r)$ at the horizon, $\phi(r_h)$, in dependence on $\psi_+$ for two different
values of $A'(r_h)$. Interestingly, we observe a new phenomenon~:
for some values of $A'(r_h)$ not too small (e.g. $A'(r_h)=-0.35$) we were able to construct two branches
of solutions. When decreasing $\psi_+$ the branches merge into a single solution at a critical
value of $\psi_+ > 0$. This means that no solution without scalar hair exists for these choices
of $A'(r_h)$ when the black string is rotating.  

For smaller values of $A'(r_h)$ (e.g. $A'(r_h)=-0.5$) only one branch of solutions is present
which continuously extends to the $\psi_+=0$ limit. In this case solutions without scalar hair
do exist.

\subsection{Holographic superfluids}
 \begin{figure}[ht]
\hbox to\linewidth{\hss%
	\resizebox{9cm}{7cm}{\includegraphics{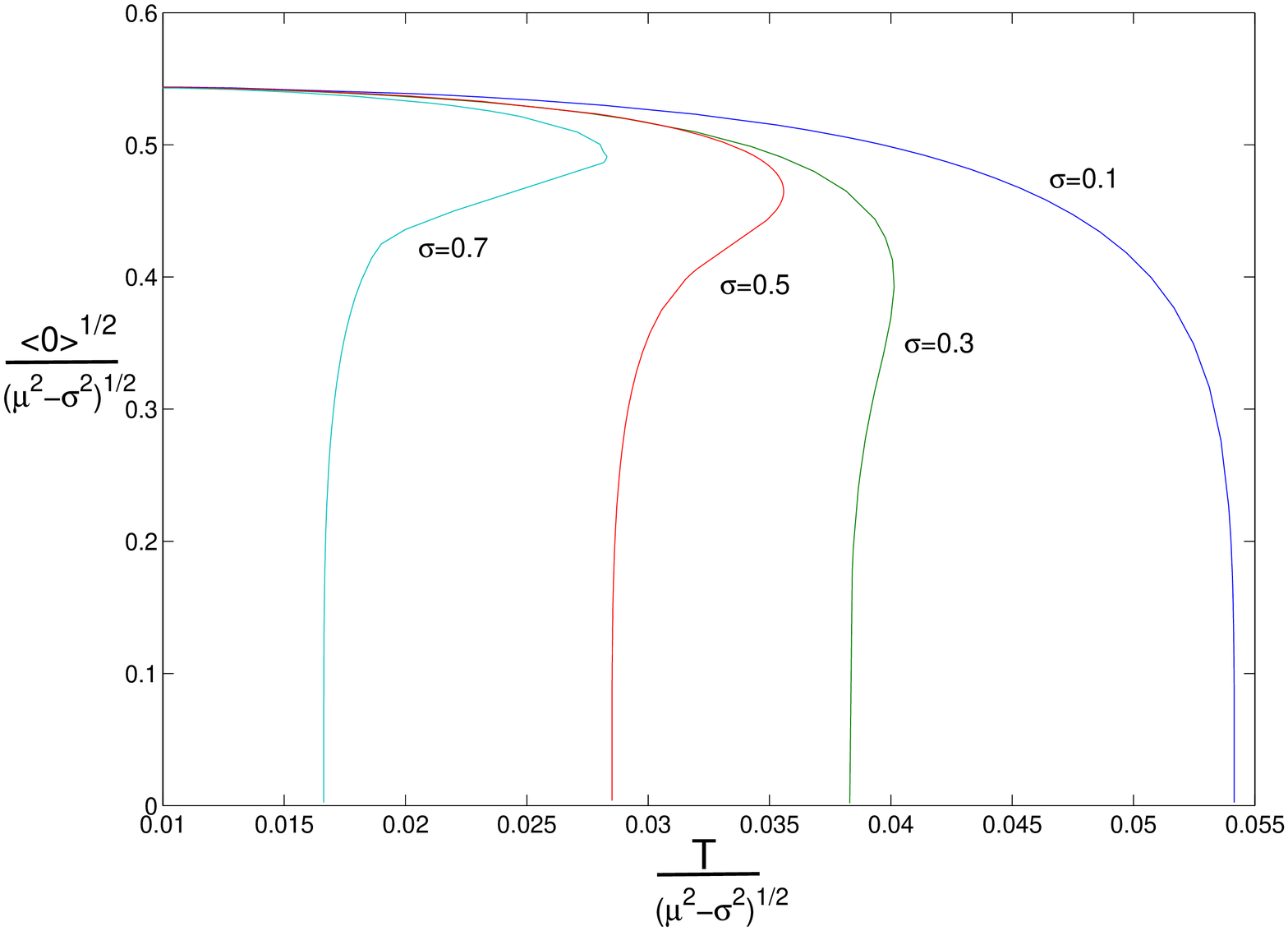}}
\hspace{5mm}%
        \resizebox{9.2cm}{6.6cm}{\includegraphics{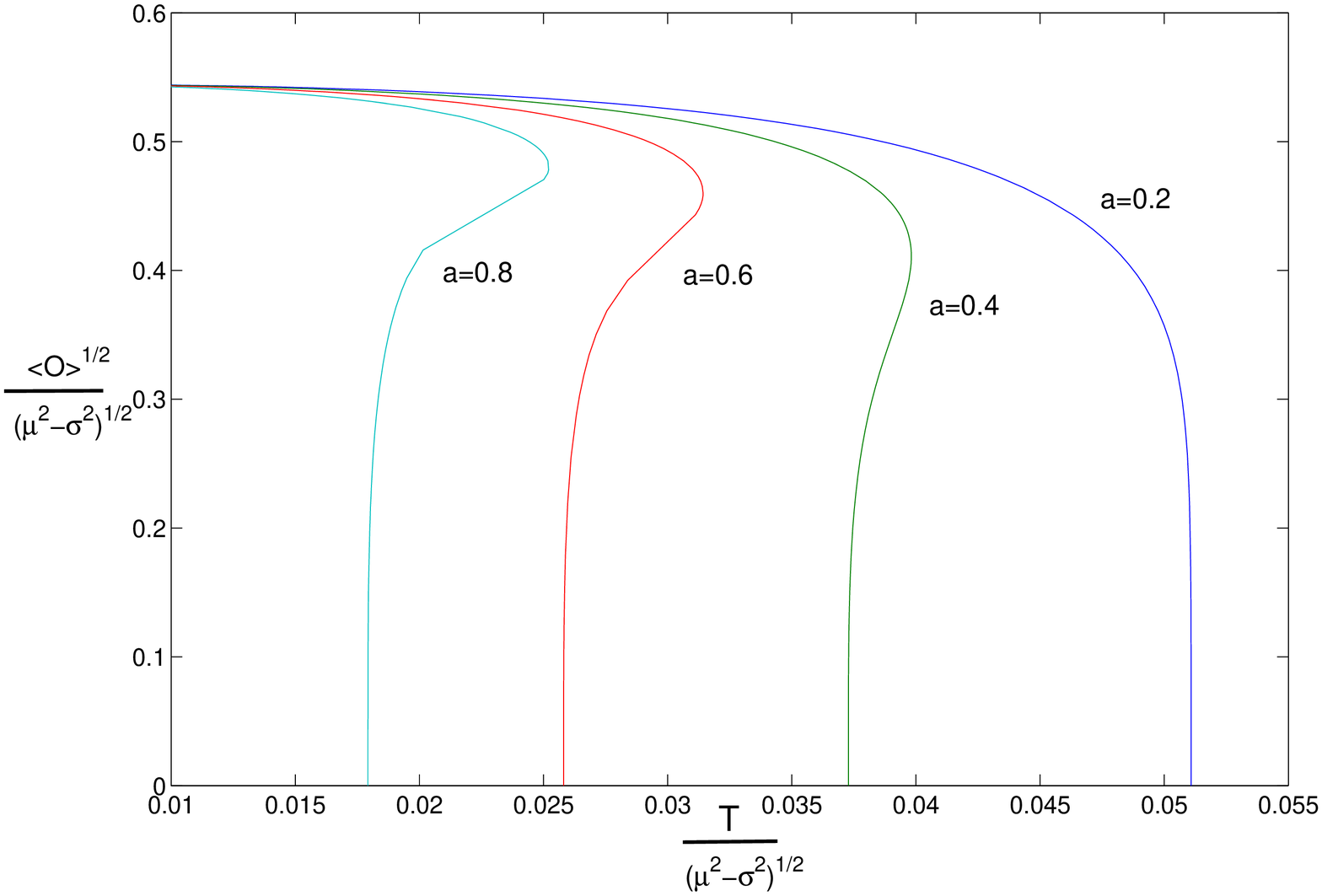}}	
\hss}
 
\caption{
{\small The condensate $<{\cal O}>^{1/2}\equiv\psi_+^{1/2}$ as function of the temperature $T$  
for $\mu=1$, $a=0.1$ and different values of the superfluid velocity $\sigma$ (left) and
for $\mu=1$, $\sigma=0.05$ and different values of the rotation parameter $a$ 
(right).
}
 }
\label{fig3}
\end{figure}
 \begin{figure}[ht]
 \hbox to\linewidth{\hss%
	\resizebox{11cm}{8cm}{\includegraphics{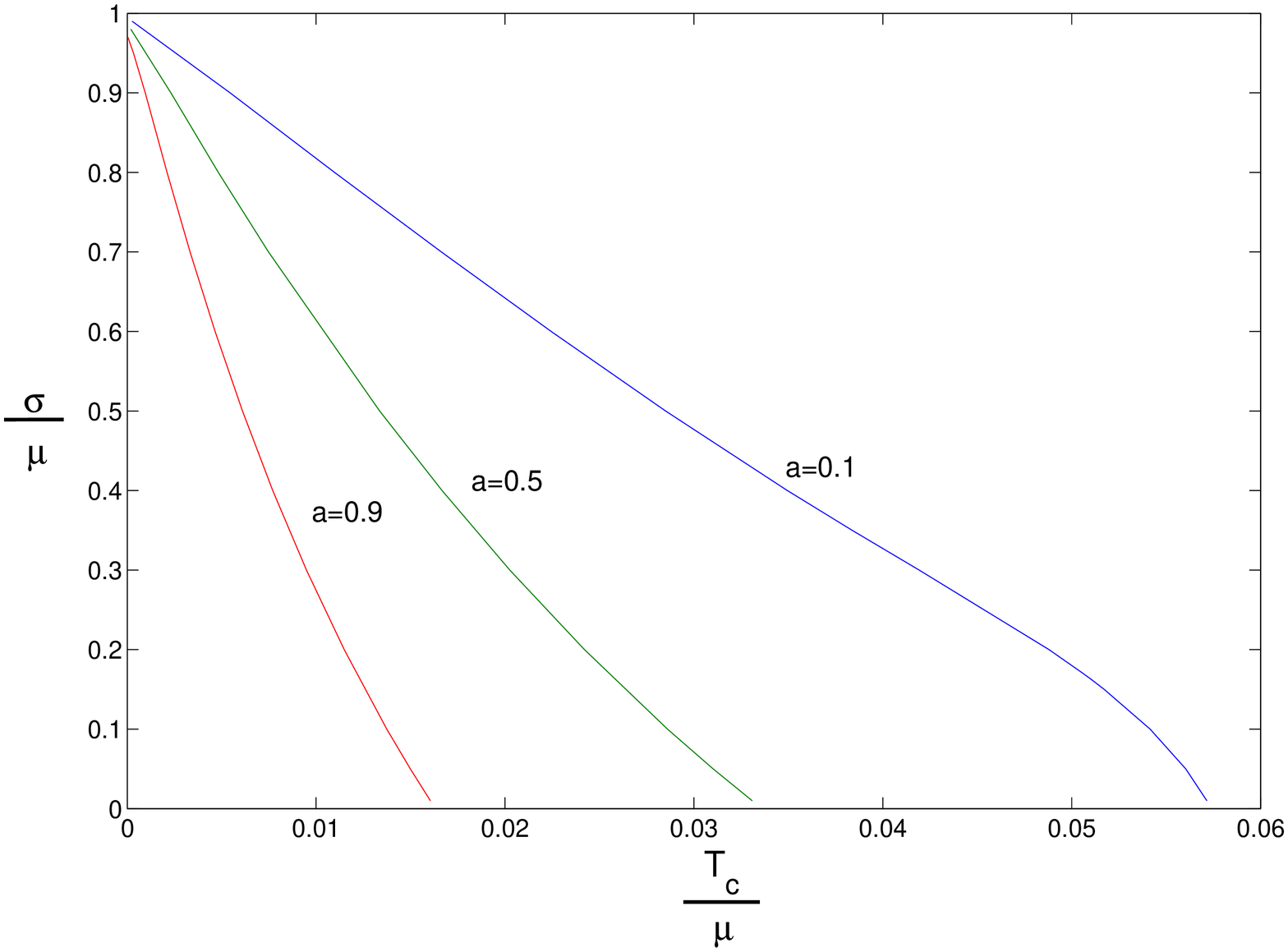}}
\hss}
\caption{
{\small The dependence of the critical temperature $T_c$ on the ratio $\sigma/\mu$ of the superfluid
velocity $\sigma$ and the chemical potential $\mu$ is shown for different values of the rotation parameter $a$. 
}
 }
\label{fig4}
\end{figure}
 \begin{figure}[ht]
\hbox to\linewidth{\hss%
	\resizebox{11cm}{8cm}{\includegraphics{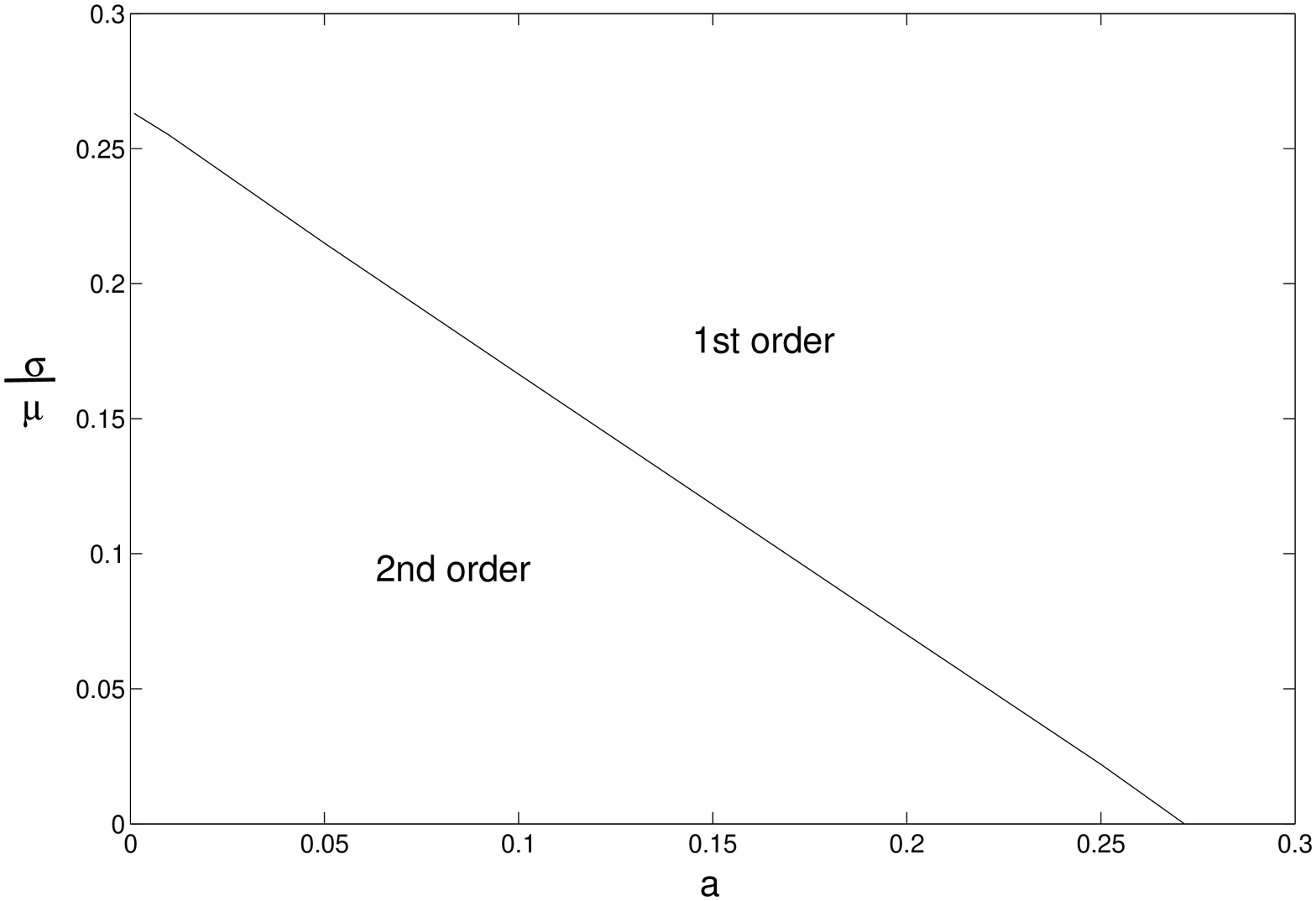}}
\hss}
\caption{{\small The value of $\sigma/\mu$ at which the phase transition changes its order (denoted by $(\sigma/\mu)_{\rm cr}$) is given as function of $a$.
Obviously for $a \ge 0.271$ the phase
transition is 1st order independent of the choice of $\sigma/\mu$. Note that the value
of $(\sigma/\mu)_{\rm cr}$ for $a=0$ is in very good agreement with that found in \cite{hks}.
}
 }
\label{fig5}
\end{figure}

We will now discuss the boundary theory of our model that lives on $\mathbb{R}^2\times S^1$.
In the following we want to work in the grand canonical ensemble, i.e. we will present our results
for fixed values of the chemical potential $\mu$ and the superfluid velocity $\sigma$.
Note that due to the scale invariance of our theory, we can always choose $\mu\equiv 1$.
For $a=0$ our model reduces to that studied in \cite{hks} with the exception that the
boundary theory in \cite{hks} lives on $\mathbb{R}^3$. It is important to note that the study
here differs from that in the gravity picture. When discussing the dual theory we allow
the horizon radius $r_h$ and hence the temperature to vary, while in the gravity picture we are interested
in the formation of scalar hair on black strings with fixed horizon radius $r_h$. For all
our numerical studies we have chosen $\ell=1$.  

First, we have studied the dependence of the value of the condensate $<{\cal O}>^{1/2}\equiv \psi_+^{1/2}$ on the temperature. Our results are shown in Fig.\ref{fig3} for fixed
$a=0.1$ and different values of $\sigma$ (left) and fixed $\sigma=0.05$ and different values
of $a$ (right), respectively. We observe that the critical temperature $T_c$ at which condensation
sets in decreases with both $a$ and $\sigma/\mu$. We show the dependence of $T_c$ on the ratio $\sigma/\mu$ of the superfluid
velocity $\sigma$ and the chemical potential $\mu$ in Fig.\ref{fig4}. Note that we did not observe
superfluidity to be destroyed, i.e. $T_c\rightarrow 0$ for any choice of $a$ and $\sigma/\mu < 1$.

We also find that the order of the phase transition changes when increasing
either the ratio $\sigma/\mu$ or $a$. While for $\sigma/\mu < (\sigma/\mu)_{\rm cr}$ the phase transition is 2nd order with a continuous
increase of the condensate with decreasing temperature $T$, the phase transition is 1st order
for $\sigma/\mu > (\sigma/\mu)_{\rm cr}$ such that the condensate increases first with increasing
$T$ up to a maximal value of the temperature and then increases further with then decreasing $T$.
This is clearly seen in Fig.\ref{fig3} where the phase transition is 2nd order for $(a,\sigma)=(0.1,0.1)$ and
$(a,\sigma)=(0.2,0.05)$, while it is 1st order for all other choices of $a$ and $\sigma$ that we present here.
Our results for $(\sigma/\mu)_{\rm cr}$ are given in Fig.\ref{fig5} where the line divides the $a-\frac{\sigma}{\mu}$--plane into two parts: one in which the phase transition is 2nd order and one
in which the phase transition is 1st order. For $a=0$ we find the value given in \cite{hks}: $(\sigma/\mu)_{\rm cr}=0.274$.
For increasing $a$ the value of the ratio $\sigma/\mu$ at which the order changes decreases, e.g.
we find for $a=0.1$ that $(\sigma/\mu)_{\rm cr}=0.167$.
Interestingly, we observe that for $a>0.271$ the phase transition is always 1st order independent of
the choice of $\sigma/\mu$. 

In order to confirm that our interpretation of the nature of the phase transition is correct we have also computed the free energy
of the solutions. Our results for $\sigma=0.1$ are given in Fig.\ref{fignew11}.
For small rotation parameter (here $a=0.1$) the superconducting phase has
always smaller free energy than the normal phase and is hence preferred. At the critical
temperature the free energy is differentiable which signals a 2nd order phase transition.
For larger rotation parameter (here $a=0.4$) there exists a metastable superconducting phase with larger free energy than the normal phase. For $T<\tilde{T}$ the free energy of the superconducting phase becomes
smaller than that of the normal phase. Moreover, the free
energy is not differentiable at the maximal temperature $T_{\rm max}$ up to which a superconducting phase
exists which means that the
phase transition is 1st order. This exactly confirms our interpretation when studying the
value of the condensate as function of temperature. 
To see how the picture changes for larger values of $\sigma$, we have also studied
the case $\sigma=0.6$. Our results are given in Fig.\ref{fignew22} for
two different values of $a$. In both cases, the free energy is not differentiable at the
maximal temperature and hence the phase transition is 1st order. Moreover, we observe
that the free energy of the metastable phase increases considerably with both $\sigma$ and $a$.
The extension of the metastable phase in $T$ compared to the maximal temperature $T_{\rm max}$ also increases with increasing $a$. For $\sigma=0.6$ and
$a=0.1$ the temperature range in which the superconducting phase has
more free energy than the normal phase is roughly $T/T_{\rm max}\in [0.7:1]$, while 
for $\sigma=0.6$ and
$a=0.4$ this is $T/T_{\rm max}\in [0.5:1]$.

 \begin{figure}[ht]
\hbox to\linewidth{\hss%
	\resizebox{11cm}{8cm}{\includegraphics{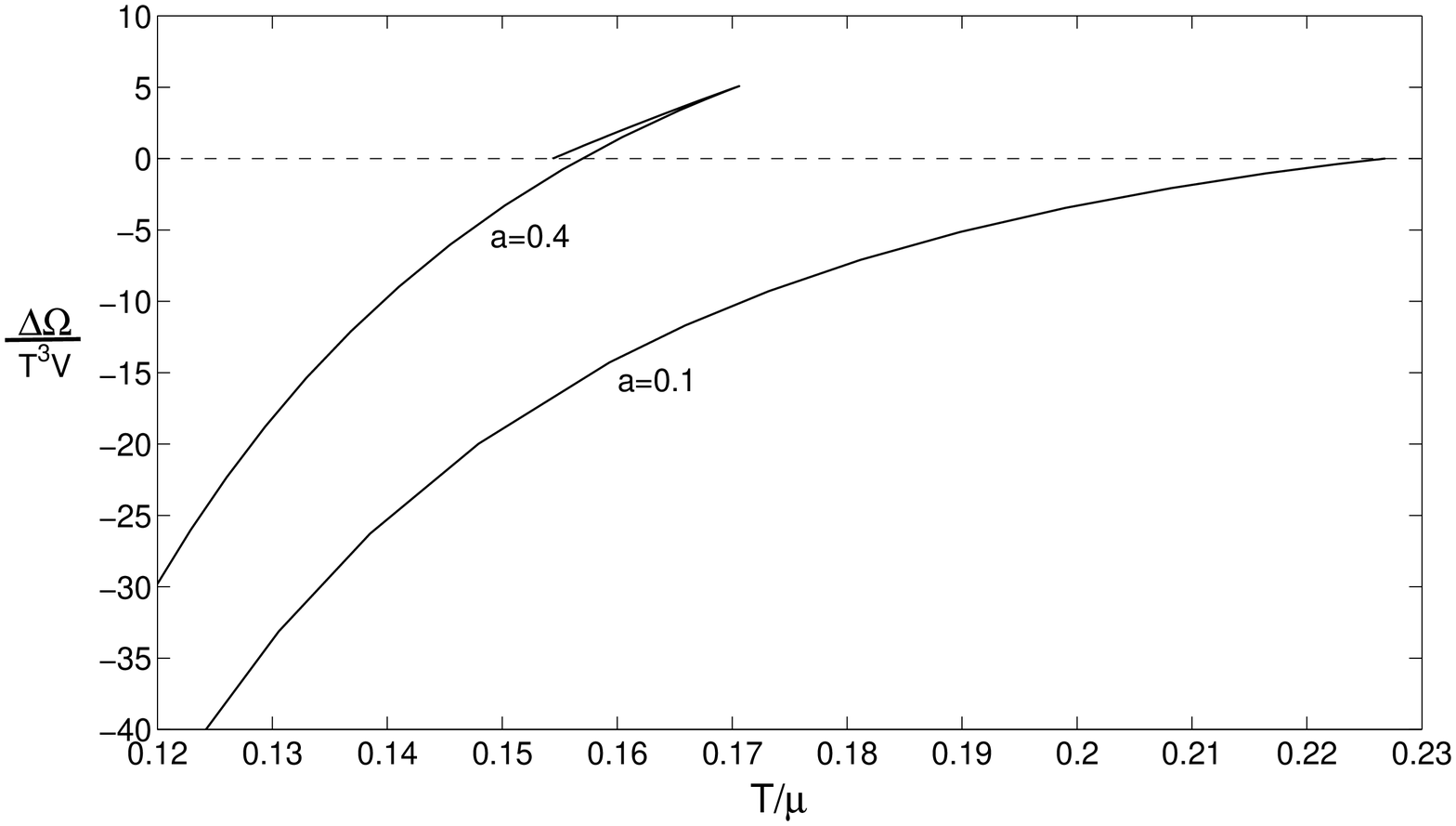}}
\hss}
\caption{{\small The value of $\Delta\Omega/(T^3 V)$ which denotes the
difference between the free energy of the superconducting phase
and the normal phase is shown for $\sigma=0.1$ and two different
values of $a$.
}
 }
\label{fignew11}
\end{figure}
 \begin{figure}[ht]
\hbox to\linewidth{\hss%
	\resizebox{11cm}{8cm}{\includegraphics{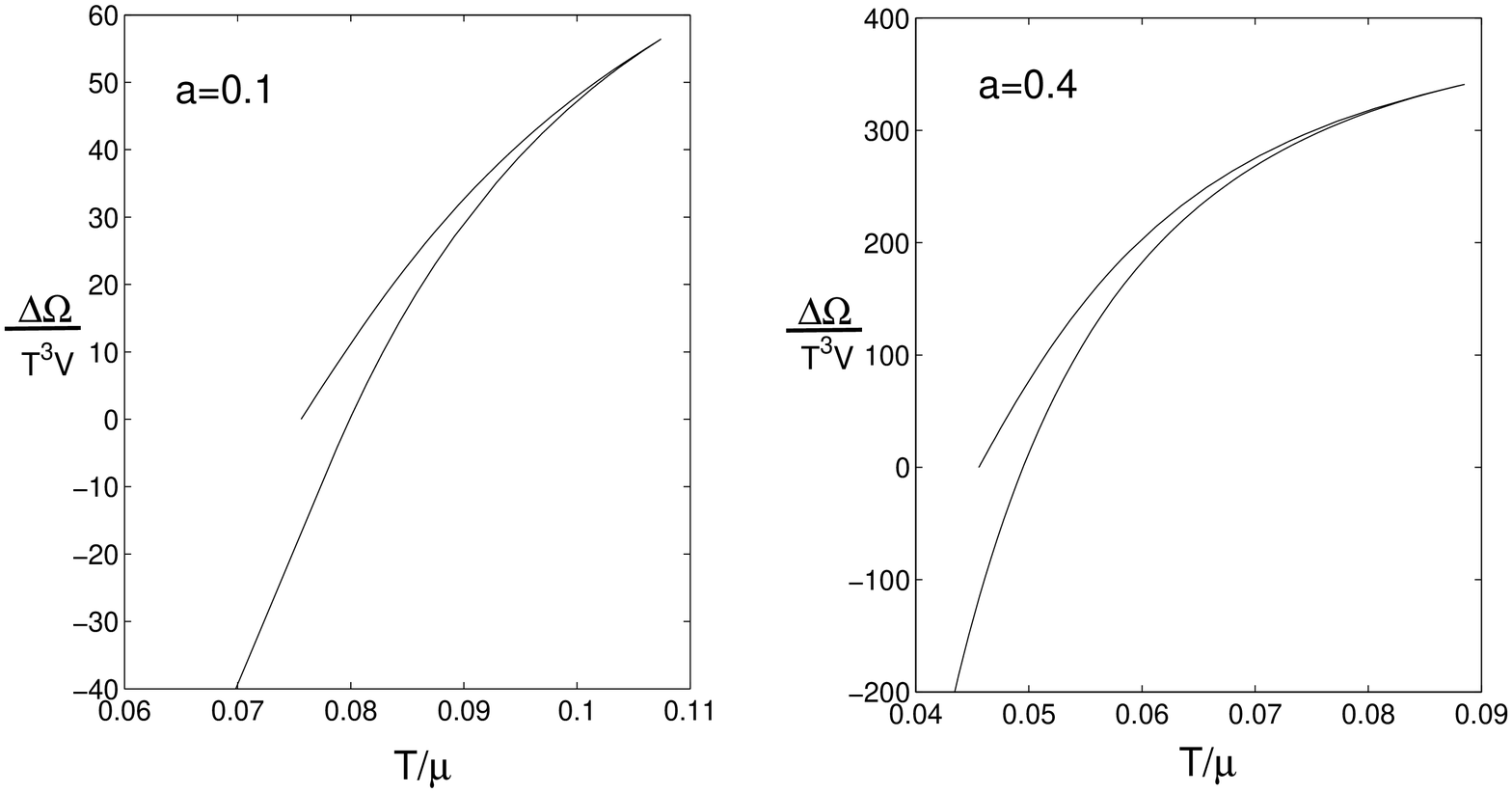}}
\hss}
\caption{{\small The value of $\Delta\Omega/(T^3 V)$ which denotes the
difference between the free energy of the superconducting phase
and the normal phase is shown for $\sigma=0.6$ and $a=0.1$ (left) and
$a=0.4$ (right), respectively.
}
 }
\label{fignew22}
\end{figure}

\section{Conclusions}
In this paper we have studied the formation of scalar hair on uncharged rotating black strings in AdS space--time.
We observe that the black strings can be dressed by scalar hair when considering a  $m^2\psi^2$-potential.
This provides a further counterexample to the No--hair conjecture.
Interestingly, we find that for some choices of the magnetic field on the horizon rotating
black strings must carry scalar hair, i.e. uncharged rotating black strings without scalar hair do not exist.

From the viewpoint of the dual theory our solutions describe (2+1)--dimensional holographic superfluids
with non--vanishing superfluid velocity. Our model generalizes the work of \cite{hks} in the sense
that we have the rotation parameter $a$ of the black string as additional parameter and that
our boundary theory lives on a space with one compact dimension. We find in analogy to \cite{hks}
that the phase transition becomes 1st order above a critical value of the
ratio between superfluid velocity $\sigma$ and chemical potential $\mu$. We observe that the critical
value decreases with increasing rotation of the black string and in fact becomes
zero for sufficiently large rotation parameter $a$ such that the phase transition is always
1st order. 

It would also be interesting to look at the formation of scalar hair on charged rotating
black strings, i.e. study the model of \cite{lemos3,dehghani3} with a different potential. We believe, however, that the results will be qualitatively similar to those presented in this paper.

We have worked in the probe limit in this paper. The investigation of the back--reaction of the space--time on the solution is left as future work. With view to the observation
that for $a=0$ the phase transition is always 2nd order in the strong back--reaction regime \cite{sonnerwithers} it will be interesting to see whether the
phase transition for $a\neq 0$ is 1st or 2nd order in this regime.

\medskip
\medskip

\noindent
{\bf\large Acknowledgments} 
YB would like to thank the Belgian FNRS for financial support.

\newpage

\begin{small}

\end{small}

\end{document}